\documentclass[%
reprint,
superscriptaddress,
nofootinbib,
 amsmath,amssymb,
 aps,
 prl,
floatfix
]{revtex4-1}

\usepackage{graphicx}
\usepackage{dcolumn}
\usepackage{bm}
\usepackage{paralist}
\usepackage[usenames,dvipsnames]{color}
\usepackage{blindtext}
\usepackage[separate-uncertainty = true,multi-part-units=single]{siunitx}
\usepackage{suffix}
\usepackage{siunitx}

\newcommand{\homo}{\mathrm{HOMO}}

\newcommand{\lmax}{L_\mathrm{max}}

\include{common_definitions}

\begin{document}

\title{Multiple orbital effects in laser-induced electron diffraction of aligned molecules}

\author{Faruk Kre\v{c}ini\'c}
\affiliation{Max Born Institute, Max-Born-Str.\ 2A, 12489 Berlin}
\affiliation{Fritz Haber Institute of the Max Planck Association, Faradayweg 4-6, 14195 Berlin}
\author{Philipp Wopperer}
\author{Biagio Frusteri}
\affiliation{Nano-Bio Spectroscopy Group and ETSF,  Dpto. Fisica de Materiales, Universidad del Pa{\'i}s Vasco,  CSIC-UPV/EHU-MPC, 20018 San Sebasti{\'a}n, Spain}
\author{Felix Brau{\ss}e}
\affiliation{Max Born Institute, Max-Born-Str.\ 2A, 12489 Berlin}
\author{Jean-Gabriel Brisset}
\affiliation{Max Born Institute, Max-Born-Str.\ 2A, 12489 Berlin}
\affiliation{Universit\'e de Gen\`eve, GAP, 22 chemin de Pinchat, 1211 Gen\`eve 4, Switzerland}
\author{Umberto De Giovannini}
\affiliation{Nano-Bio Spectroscopy Group and ETSF,  Dpto. Fisica de Materiales, Universidad del Pa{\'i}s Vasco,  CSIC-UPV/EHU-MPC, 20018 San Sebasti{\'a}n, Spain}
\author{Angel Rubio}
\affiliation{Nano-Bio Spectroscopy Group and ETSF,  Dpto. Fisica de Materiales, Universidad del Pa{\'i}s Vasco,  CSIC-UPV/EHU-MPC, 20018 San Sebasti{\'a}n, Spain}
\affiliation{Max Planck Institute for the Structure and Dynamics of Matter, Luruper Chaussee
149, 22761 Hamburg, Germany}
\author{Arnaud Rouz\'ee}
\email{rouzee@mbi-berlin.de}
\author{Marc J.J. Vrakking}
\affiliation{Max Born Institute, Max-Born-Str.\ 2A, 12489 Berlin}




\date{\today}

\begin{abstract}
Photoelectron Angular Distributions (PADs) resulting from 800 nm and 1300 nm strong field ionization of impulsively aligned CF$_3$I molecules were analyzed using time-dependent density functional theory (TDDFT). The normalized difference between the PADs for aligned and anti-aligned molecules displays large modulations in the high-energy re-collision plateau that are assigned to the diffraction of back-scattered photoelectrons. The TDDFT calculations reveal that, in spite of their 2.6 eV  energy difference, ionization from the HOMO-1 orbital contributes to the diffraction pattern on the same footing as ionization from the doubly degenerate HOMO orbital.  
\end{abstract}

\maketitle

Following structural changes within single molecules on their natural time and length scales is one of the great challenges in ultrafast molecular physics. Large efforts are currently devoted to the development of techniques for the direct imaging of nuclear motion with atomic resolution. 
Diffractive imaging methods using ultrashort X-ray pulses available at Free Electron Lasers \cite{Emma,Ackermann}, or using ultrashort electron pulses \cite{uelec:Dwyer, uelec:Zewail,uelec:Yang2016}, have the potential to record structural information with the spatio-temporal resolution required for obtaining "molecular movies" \cite{uelec:Dwyer,txtbook:Zewail,fel:Glownia,uelec:Yang2016}. In both approaches however, realizing single molecule imaging with sub-10 fs temporal resolution has proven challenging \cite{Schulz2015,Gliserin2015}, since the required synchronization between the visible/ultra-violet laser pulses initiating the molecular dynamics of interest and the X-ray/UED probe is difficult to achieve.

Fully laser-based molecular self-imaging techniques using strong field ionization by an intense infrared (IR) laser pulse are an alternative and promising route towards the imaging of (time-dependent) molecular structures in the gas phase \cite{Lein2007}.
In particular, Laser-Induced Electron Diffraction (LIED) \cite{lied:Niikura,lied:Lein,lied:Spanner,lied:Meckel}, where the ionization of a molecule by a strong IR laser field leads to the creation of a photoelectron wavepacket that is accelerated by the laser field to induce a recollision with the parent molecular ion, has already demonstrated few-femtosecond and sub-\AA ngstr\"{o}m resolution \cite{lied:Blaga,lied:Pullen,lied:Wolter}. The time resolution in LIED is given by the optical cycle of the driving laser field  \cite{lied:Blaga,lied:Wolter} and can reach the sub-femtosecond timescale, whereas high spatial resolution is possible due to the high kinetic energy of the re-colliding photoelectron, which determines its De Broglie wavelength and can reach values of 0.1 \AA\ when using mid-infrared laser fields.

Retrieval of the molecular structure from an LIED experiment is often done in the framework of the Quantitative Rescattering Theory (QRT) \cite{lied:Spanner,qrs:Chen,lied:Yu}, which usually assumes that (i) the ionization takes place from the Highest Occupied Molecular Orbital (HOMO) and that (ii) the initial shape of the electron wavepacket is lost during its propagation in the oscillatory laser field, so that the re-colliding electron wavepacket can be approximated by a plane wave. Both of these assumptions may be questioned. Strong field ionization, in particular of polyatomic molecules, often involves multiple ionization pathways corresponding to the removal of electrons from different orbitals \cite{Boguslavskiy}, as shown by previous high harmonic generation experiments in aligned CO$_2$  \cite{Smirnova} and strong field ionization experiments performed in hydrocarbons \cite{Boguslavskiy}, whereas recent work from our laboratory has explicitly demonstrated the breakdown of the plane wave approximation \cite{sfa:Schell}. 

In this letter, we study laser-induced electron diffraction of CF$_3$I. Photoelectron Angular Distributions (PADs) of impulsively aligned and strong-field ionized CF$_3$I molecules were recorded for different alignment distributions, laser wavelengths and intensities. By comparing our experimental data with ab-initio calculations using TDDFT, we conclude that the PADs contain contributions that can be assigned to the two highest occupied molecular orbitals, i.e.\ the HOMO and HOMO-1. 
Our experimental and theoretical investigations indicate that in polyatomic molecules an accurate description of LIED requires the inclusion of multiple ionization channels. 

\begin{figure}
\centering
\includegraphics[scale=0.4]{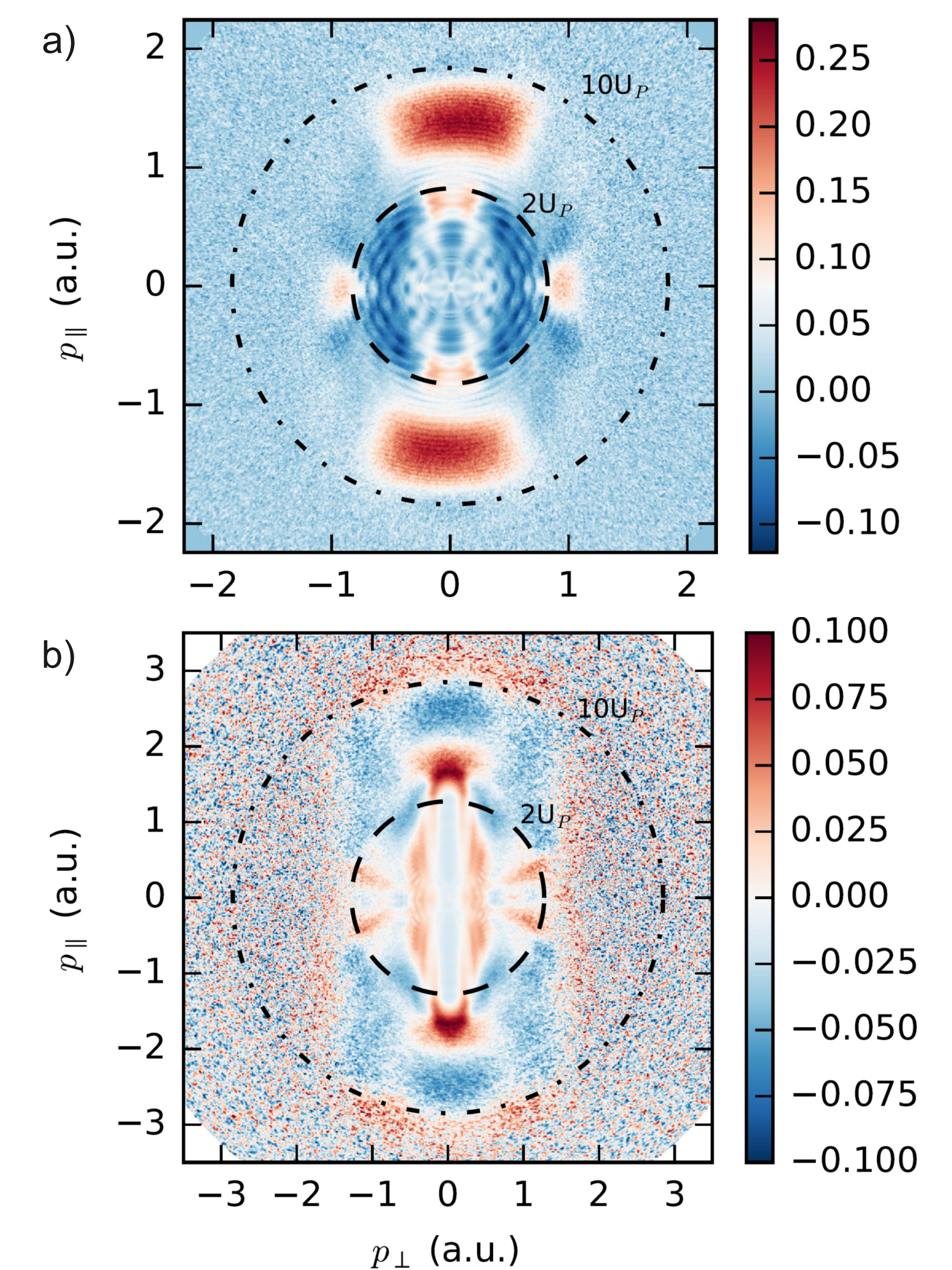}
\caption{Experimental normalized differences of photoelectron momentum distributions $I_\Delta$ (see Eqn. (1)) recorded for aligned and anti-aligned CF$_3$I for (a) a probe laser intensity of \SI{7.5(8)e13}{W/cm^2} at a central wavelength of 800 nm ($U_p$ = 4.4 eV) and (b) a probe laser intensity of \SI{7.0(6)e13}{W/cm^2} at a central wavelength of 1300 nm ($U_p$ = 11.1 eV). Iso-energy lines are indicated at $2U_p$ (dashed line) and $10U_p$ (dash-dotted line).}
\label{fig:800PAD}
\end{figure}
In our experiment, a Ti:Sapphire laser system was used delivering 2 mJ, 30 fs pulses at a 1 kHz repetition rate. The output of the laser was split into two pulses. One 800 nm pulse (1 mJ) was stretched to a 1.2 ps pulse duration in a 10 cm long SF11 glass block, in order to enable dynamic alignment of the CF$_3$I molecules \cite{Vrakking}. Within the current experiment we consider this the pump laser pulse. The second 800 nm laser pulse was either directly used as probe laser, or was used to pump an optical parametric amplifier (TOPAS-C from Light Conversion) in order to generate \SI{150}{\micro J}, 1300 nm probe laser pulses. Both co-polarized laser pulses were focused inside a Velocity Map Imaging Spectrometer (VMIS) \cite{VMI:Eppink} using a 20 cm lens. 
The waist of the alignment pulse was adjusted with a telescope in order to suppress ionization from the pump pulse while ensuring the highest degree of alignment. 
The probe intensity was adjusted using a $\lambda/2$-waveplate and a polarizer, resulting in an intensity in the interaction region between \SI{4e13} and \SI{2e14}{W/cm^2}. 
At the center of the VMIS, the laser pulses interacted with a cold molecular beam of CF$_3$I seeded in helium (seed ratio 1:100) produced by an Even-Lavie valve running at 500 Hz \cite{mbeam:Even}. Charged particles produced by the probe laser were detected by a microchannel plate/phosphor screen assembly and recorded with a CCD camera. 
The VMIS was used to record 2D projections of the photoelectron momentum distributions that, exploiting the cylindrical symmetry in the experiment, were used to extract initial 3D photoelectron momentum distributions using the BASEX method \cite{VMI:Dribinski}. 

The 1.2 ps long alignment pump pulse, which was polarized parallel to the plane of the detector, was used to populate a rotational wavepacket. The field-free evolution of the wavepacket led to \emph{alignment revivals} \cite{align:Seideman, Vrakking} at regular intervals given by t=$n\tau_r/2=n\times163.7$ ps ($\tau_r=1/(2cB)$= 327.4 ps and $B=0.0509$ cm$^{-1}$ for CF$_3$I), where n is an integer. The degree of alignment was characterized by recording the I$^+$ momentum distribution resulting from Coulomb explosion of the molecules by the 800 nm probe laser \cite{fel:Kupper,Vrakking}, which in this case was polarized perpendicular to the detector plane, i.e. perpendicular to the alignment laser polarization. 
An estimate of $\left\langle \cos^2 \theta \right\rangle$, with $\theta$ the angle between the laser polarization axis and the molecular axis, was obtained by comparing the measured pump-probe delay-dependent fragment angular distributions to a prediction solving the time-dependent Schr\"{o}dinger equation, where the pump intensity and temperature were used as fitting parameters. 
Best agreement was found for a sample temperature of 7 K and a peak pump laser intensity of \SI{5e12}{W/cm^2}, leading to a degree of alignment $\left\langle \cos^2 \theta \right\rangle$ of 0.75.

Normalized differences of photoelectron angular distributions (PADs) recorded for aligned and anti-aligned molecules given by
\begin{equation}
\label{eq:aminbnorm}
I_{\Delta}=\frac{I_{align}-I_{anti-align}}{I_{align}+I_{anti-align}}
\end{equation} 
are shown in Fig. \ref{fig:800PAD}(a-b).   
Large modulations are observed in the angular distribution throughout the normalized difference maps at both probe laser wavelengths. In the semi-classical picture of strong field ionization, the final electron momentum distribution can be divided into two regions \cite{sfa:Corkum}. 
Photoelectrons that do not interact with the molecular ion after ionization (direct electrons) can classically reach final energies up to 2$U_p$, where $U_p=I/4\omega^2$ (in a.u.) is the ponderomotive energy, whereas photoelectrons that have re-collided with the parent ion may reach an energy up to 10$U_p$. 
In our experiment, the direct electron contribution drops off rapidly above the 2$U_p$ cut-off, which is indicated by dashed circles in Fig. \ref{fig:800PAD} (a-b).
In the low-momentum region, direct \emph{and} re-scattered photoelectrons interfere with each other and both contribute to the observed modulations in the normalized difference images \cite{eholo:Huismans}. In the re-collision plateau, i.e.\ between 2$U_p$ and 10$U_p$, a pronounced modulation of the normalized difference is observed along the laser polarization axis ($p_{\parallel}$), both in the 800 nm and the 1300 nm data. 
The 800 nm data in Fig. \ref{fig:800PAD} (a) shows pronounced positive values in the normalized difference along the $p_{\parallel}$-axis for momenta $> 1$ a.u.
In the 1300 nm data shown in Fig. \ref{fig:800PAD} (b), positive values are visible for $p_{\parallel}$ near 1.7 a.u., followed by negative values for momenta $> 2$ a.u.
In the 2$U_p$ to 10$U_p$ energy range, the contribution of back-scattered photoelectrons that have experienced a hard re-collision with the parent ion are dominant \cite{lied:Xu}.
Following strong-field ionization, the maximum photoelectron re-collision energy is given by 3.17 $U_p$ \cite{sfa:Corkum,SchaferPRL}, corresponding to a De Broglie wavelength of 3.3 and 2.0 \AA for the 800 nm and 1300 nm laser wavelengths, respectively, at an intensity of \SI{7.5e13}{W/cm^2}. 
Considering that the I--F and C--I internuclear distances are 2.7 and 2.14 \AA, respectively, we expect that the observed trends in the normalized difference images in the re-collision plateau are a consequence of the molecular structure.

Ideally, we would like to analyze the experimental results using the common approach based on the QRT  (see Refs.\ \cite{lied:Spanner,qrs:Chen,lied:Yu}), where the photoelectron momentum distribution in the re-collision plateau is described as the product of a momentum distribution of the returning electron wave packet $W (p_r)$ and a differential-scattering cross section describing a (field-free) collision of the laser-driven re-collision electron with the target ion. The former is then evaluated using either an effective atomic ADK rate \cite{sfa:Ammosov}, or the MO-ADK tunneling model introduced by Tong \textit{et al.} \cite{msfa:Tong,Zhang2015}. However, neither approach turned out to be successful in the case of CF$_3$I, given the fact that atomic ADK does not contain any angular dependence, while MO-ADK underestimates contributions to the ionization from lower-lying orbitals. The latter is usually calculated using an independent atom model (IAM), which is likely to fail given the modest kinetic energies of the recolliding electrons in our experiment.

As an alternative to the QRT method, we performed ab-initio calculations using TDDFT. The (adiabatic) local-density approximation (ALDA) \cite{PW92} was used, with an average-density self-interaction correction (SIC) \cite{Leg02} which corrects the tail of the Coulomb potential and yields an accurate ionization potential. A Cartesian grid with a maximum range of $80\,a_0$ in all three directions was chosen, which was large enough to accommodate the quiver length $\alpha$ of the photoelectrons in the experiment ($\alpha=F/\omega^{2}=43.51 a_{0}$ for a 1300 nm laser field with a peak intensity of I=\SI{1e14}{W/cm^2}) . The ionic cores were described by norm-conserving pseudo-potentials. In this configuration, the calculated single-particle energies of the two highest occupied molecular orbitals were $\epsilon_\homo=-10.2$\,eV and $\epsilon_{\homo-1}=-12.6\,$eV, in good agreement with experimental values \cite{Bancroft}. 
Photoelectron spectra were calculated with the time-dependent surface flux method \cite{Wopperer} using a spherical surface located at $r=50\,a_0$. Angular momenta up to $\lmax = 40$ were included. Ejected electrons were absorbed by a complex absorbing potential~\cite{DeGiovannini:2015} with a width of $L=30\,a_0$ and a height of $\eta = -0.2$\,a.u.\ located at $r>50\,a_0$. The TDDFT calculations yielded three-dimensional PADs for fixed orientations of the molecule with respect to the alignment laser pulse polarization. In order to compare to the experimental results, weighted averages of the simulated PADs were constructed using the alignment/anti-alignment distributions extracted from the experiment.

\begin{figure}
\centering
\includegraphics[scale=0.45]{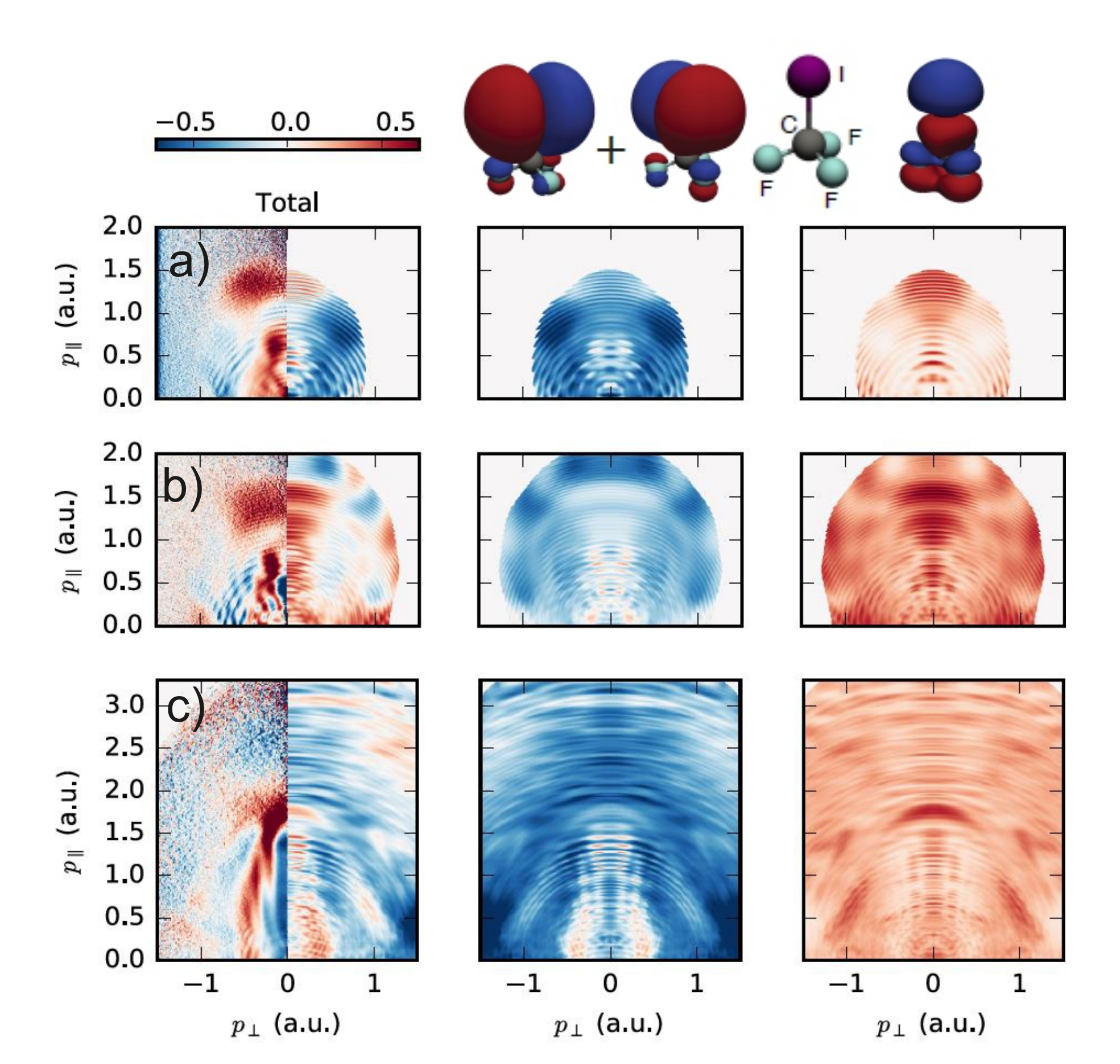}
\caption{Comparison between experimentally measured normalized difference maps and TDDFT calculations for different probe laser configurations: (a) 800 nm and \SI{3.4e13}{W/cm^2}, (b) 800 nm and \SI{7.7e13}{W/cm^2}, and (c) 1300 nm and \SI{7.0e13}{W/cm^2}. The first column compares the experiment (left half) with the total result of the TDDFT calculation (right half). The second and third columns correspond to the contribution to the difference maps from the HOMO and HOMO-1 orbitals, respectively (i.e.\ a separation of $I_{\Delta}$ into contributions from the HOMO and HOMO-1, in both cases normalized to the total HOMO + HOMO-1 ion yield). The HOMO and HOMO-1 Kohn-Sham orbitals are displayed at the top of the figure.}
\label{fig:exp-vs-tddft}
\end{figure}

In Fig. \ref{fig:exp-vs-tddft} (first column) we show a comparison between the experimental normalized difference momentum maps and the TDDFT simulations for three different laser intensities and two wavelengths.
In all the simulated cases, the TDDFT results show significant levels of agreement with the experiments. In particular, the positive feature appearing at $p_\parallel\approx 1.4\,$a.u.\ for $\lambda=800\,$nm (Fig.~\ref{fig:exp-vs-tddft} (a) and (b)) and, to a lesser extent, at $p_\parallel\approx 1.7\,$a.u.\ for $\lambda=1300\,$nm (Fig.~\ref{fig:exp-vs-tddft} (c)) match quite nicely. This supports the suitability of TDDFT for describing the dynamics of the LIED process. We note that remaining differences between the experimental and TDDFT results are attributed to the role of focal volume averaging (which is not included in the calculations) in the experiment.

An attractive feature of TDDFT is that it permits extraction of the contributions of each Kohn-Sham orbital to the final spectra. We find that the two highest occupied orbitals, i.e.\ the HOMO and the HOMO-1, contribute significantly to the normalized difference image, whereas the ionization from more strongly bound states is negligible. The contributions of the HOMO and the HOMO-1 to the normalized difference image are shown in the second and third column of Fig.~\ref{fig:exp-vs-tddft}. Remarkably, both orbitals contribute in very different ways to $I_\Delta$: the normalized difference for the HOMO (second column) is predominantly negative (i.e. dominated by anti-aligned molecules), while for the HOMO-1 (third column), the normalized difference is predominantly positive (i.e. dominated by aligned molecules). Accordingly, positive (red) features in the total normalized difference image are caused by the HOMO-1, while negative (blue) features originate from the HOMO. This behavior can be understood from the shape of the HOMO and HOMO-1 orbitals: ionization from the HOMO is suppressed for aligned molecules due to the presence of a nodal plane along the C--I bond axis (see Fig. \ref{fig:exp-vs-tddft}), whereas the HOMO-1 has perpendicular nodal planes, leading to a suppressed ionization in the anti-aligned configuration. Whereas previous experimental studies of strong field molecular ionization have hinted that multiple orbital effects may leave their imprint in LIED measurements, this analysis unambiguously shows that contributions arising from multiple orbitals are indeed present and significant.

As discussed previously, the analysis of LIED experiments can be perfomed in the framework of the QRT model, in which the PAD is expressed as the product of a recollision electron momentum distribution and a field-free differential cross-section. At sufficiently high re-collision energies, as considered here, the differential-scattering cross section describing the collision of the laser-driven photoelectron with the target ion is mainly sensitive to the interactions with the molecular charge distribution, which are nearly identical in both the HOMO and HOMO-1 ionization channels considered here. Therefore, according to the QRT model, differences in LIED patterns for the HOMO and HOMO-1 channels as observed in Fig. \ref{fig:exp-vs-tddft} must be the result of differences in the structure of the re-colliding electron wavepackets, which carry signatures of the molecular orbitals from which the photoelectrons are produced \cite{sfa:Schell}.

In conclusion, we have presented a series of experiments on LIED of aligned CF$_3$I. Using a theoretical approach based on TDDFT, contributions from multiple orbitals could clearly be identified, as a result of the fact that the orbital from which a photoelectron is removed significantly influences the PAD that can be measured following an electron-ion recollision. Therefore, models that assume ionization restricted to the highest occupied molecular orbital and considering an incoming plane wave for the returning electron wavepacket are not appropriate for describing LIED processes. One of the future applications of LIED is its application to time-resolved molecular dynamics, where a pump laser initiates a photochemical process and where an LIED measurement takes a time-resolved snapshot of the evolving molecular structure. Our study suggests that successful strategies will need to incorporate measurements of the alignment- and channel-dependent ionization rates, for example using coincidence spectroscopy \cite{lied:Pullen2016} or the CRATI technique \cite{Boguslavskiy}, in order to characterize the structure of the re-colliding photoelectron wavepacket. Under these conditions, and using sufficiently high recollision electron energies, recovery of (time-dependent) molecular structural information using an independent atom model (IAM) will be possible, permitting the recording of molecular movies.

\begin{acknowledgments}
F.K. would like to acknowledge funding  under European Community 7$^{\text{th}}$ Framework Programme under contract ITN-2008-238362 (ATTOFEL).
P.W., B.F. U.D.G. and A.R. acknowledge financial support from the European Research Council (ERC-2015-AdG-694097), Grupos Consolidados (IT578-13), and European Unions Horizon 2020 Research and Innovation program under Grant Agreements no. 676580 (NOMAD).  A. R. would like to thank the Deutsche Forschungsgemeinschaft,
Schwerpunktprogramm 1840 (SPP 1840 QUTIF project).
The authors would like to thank Dr. Frank Noack for providing access to the laser system used to perform the 1300 nm wavelength experiments. 
\end{acknowledgments}

\bibliography{CF3IPRLmainref}{}

\end{document}